\documentclass[proof]{pasj00}

\begin{document}
\SetRunningHead{Author(s) in page-head}{Running Head}

\title{The Schmidt Law at High Molecular Densities}

\author{Shinya \textsc{Komugi}, Yoshiaki \textsc{Sofue}, Hiroyuki
\textsc{Nakanishi}*, Sachiko \textsc{Onodera} \and Fumi \textsc{Egusa}}
\affil{Institute of Astronomy, University of Tokyo, 2--21--1 Osawa,
Mitaka-shi, Tokyo, 181--8588, Japan}

\affil{
*Nobeyama Radio Observatory, Minamisaku, Nagano, 384-1305, Japan}
\email{skomugi@ioa.s.u-tokyo.ac.jp}


\maketitle

\begin{abstract}
We have combined H$\alpha$ and recent high resolution $\atom{CO}{}{12}$ (J=1-0) data to consider
 the quantitative relation between gas mass and star formation rate, or
 the so-called Schmidt law in nearby spiral galaxies at regions of high
 molecular density.  The relation between gas quantity and star
 formation rate has not been previously studied for high density
 regions, but using high resolution CO
 data obtained at the NMA(Nobeyama Millimeter Array), we have found that the
 Schmidt law is valid at densities as high as $10^3 \mathrm{M_\odot} \mathrm{pc}^{-2}$
 for the sample spiral galaxies,
 which is an order of magnitude denser than what has been known to
 be the maximum density at which the empirical law holds for
 non-starburst galaxies.  Furthermore,
 we obtain a Schmidt law index of $N=1.33\pm0.09$ and roughly constant star formation
 efficiency over the entire disk, even within the
 several hundred parsecs of the nucleus.  These results imply that the
 physics of 
 star formation does not change in the central regions of spiral
 galaxies.  Comparisons with starburst
 galaxies are also given.  We find a possible discontinuity in the
 Schmidt law between normal and starburst galaxies.
\end{abstract}

\section{Introduction}

The global, disk-averaged star formation rate $\Sigma_{\mathrm{SFR}}$ and gas density $\Sigma_{\mathrm{SMD}}$
in nearby galaxies are known to be well correlated, producing a simple
power law
called the Schmidt law; $\Sigma_{\mathrm{SFR}} \propto
\Sigma_{\mathrm{SMD}}^N$ \citep{schmidt}.  The
determination of the index $N$ has been extensively studied, producing
values between 1 and 2.  However, the
density range at which this law holds for normal galaxies has not been
studied as much.  The lower density
cutoff for the Schmidt law has been argued by \citet{K89} based on
gravitational instability, giving a critical gas density of $5$ to $10$
$\mathrm{M_{\solar}} \mathrm{pc}^{-2}$ where star formation is strongly
suppressed below this limit.  On the other hand, there has been only marginal success for the
observational consideration of the Schmidt law at higher densities. This
is primarily due to the limit in spatial resolution of CO
observations, where the  $\atom{CO}{}{12}$(J=1-0) emission at
 $\lambda$=2.6mm is used as a tracer of molecular gas.

 High molecular density regions are spatially small in general, owing to
 the molecular cloud's patchy structure and central condensation \citep{sakamoto99}.  Therefore, it is
 essential to make CO observations
with smaller beam sizes (higher resolution).   \citet{Rownd99} have shown that the Schmidt law holds better when sampled within
galaxies, using 45 arcsecond resolution data at the center of over 100
nearby galaxies (see figure 11 in \citet{Rownd99}).  However, the densest gas in galaxies is normally confined to
within a few hundred parsecs of the nucleus (e.g., \citet{sakamoto99}, \citet{sofue03}).  Therefore, 45 arcsecond
resolution, which corresponds to kpc scales in nearby galaxies,
is too large to consider the validity of the Schmidt law at the highest
densities.   \citet{K98} has shown that the Schmidt law holds at higher densities
if we include a sample of IR
luminous starburst galaxies with $\mathrm{L_{FIR}} \sim 10^8 - 10^{10}
 \mathrm{L_{\solar}}$ to $\mathrm{L_{FIR}} \le 10^{12} \mathrm{L_{\solar}}$ at high molecular densities.  Including this
sample has extended the Schmidt law up to an unchallenged value of
$\Sigma_{\mathrm{SMD}} \sim  10^5 \mathrm{M_{\solar}} \mathrm{pc}^{-2}$
 (see figure 6 in \citet{K98}).  However,
 IR luminous starbursts represent special physical conditions which
 display a systematic difference from normal spirals, and including this sample in considering the validity of the
Schmidt law may lead us to a misunderstanding of the physical properties
which underly the poorly understood power law.  \citet{Gao04} have
indeed found that the IR luminous galaxies with $\mathrm{L_{IR}} \le 10^{11} \mathrm{L_{\solar}}$ do display a systematic
difference from normal galaxies in view of the Schmidt law, in that the
 ratio of HCN to CO rises abruptly for IR luminous galaxies
 compared to normal galaxies.
Exclusion of these IR luminous samples from Kennicutt(1998) and the consideration of
only normal spirals reveal that the law is only valid up to $\sim 10^2
\mathrm{M}_{\solar} \mathrm{pc}^{-2}$.   
Other studies of the Schmidt law at high densities involve HCN \citep{Gao04} and higher
transition emission such as $\atom{CO}{}{12}$ (J=2-1) as probes of
hydrogen at higher densities, but the weakness of this emission and
the difficulty in comparing $\Sigma_{\mathrm{SMD}}$
derived from these probes and others of lower density (such as CO(J=1-0)
and HI) has limited the consistent consideration of the Schmidt
law's validity at a wide density range.  

In order to approach to a more fundamental
understanding of the relation between gas content and star formation, it is essential to know to
what density the Schmidt law holds for normal galaxies, or whether the
simple power-law formulation of the Schmidt law is valid at all.  This will
require that we use a consistent probe of the gas, at a wide density range.

In this study, we consider the validity of the Schmidt Law including densities
above this limit by using recent high resolution single dish and interferometry
observations.  Section 2 provides information on CO and H$\alpha$ data,
and the procedures for calculating $\Sigma_{\mathrm{SMD}}$ and $\Sigma_{\mathrm{SFR}}$.
In section 3 we show the result, and in section 4 we discuss the various
uncertainties which relate to the Schmidt law, and the validity of the
Schmidt law at high densities.  Conclusions will be given in section 5.

\begin{table}
  \caption{Basic properties of the sample galaxies}\label{prop} 
  \begin{center}
     \begin{tabular}{cccccccc} \hline \hline
     Galaxy & R.A.& Decl.& Type & D(Mpc) & i(degrees) & size (arcmin) \\ \hline
       (1) & (2) & (3) & (4) & (5) & (6) & (7) \\  \hline     
     NGC2903 & 09 29 19.9 & +21 43 19 & SAB & 6.3 & 60 & 11.2\\
     NGC3593 & 11 11 59.2 & +13 05 28 & SA  & 5.5 & 67 & 4.9\\
     NGC4041 & 11 59 38.7 & +62 25 03 & SA  & 22.7 & 18 & 2.9\\ 
     NGC4192 & 12 11 15.4 & +15 10 23 & SAB & 16.8 & 74 & 7.8\\
     NGC4212 & 12 13 06.4 & +14 10 45 & SA  & 16.8 & 47 & 2.9\\
     NGC4254 & 12 16 16.9 & +14 41 46 & SA  & 16.8 & 28 & 5.5\\
     NGC4303 & 12 19 21.4 & +04 44 58 & SAB & 16.8 & 25 & 6.2\\
     NGC4321 & 12 20 23.2 & +16 06 00 & SAB & 16.8 & 28 & 7.1\\
     NGC4419 & 12 24 25.1 & +15 19 28 & SB  & 16.8 & 67 & 2.8\\
     NGC4501 & 12 29 28.1 & +14 41 50 & SA  & 16.8 & 58 & 6.3\\
     NGC4535 & 12 31 47.9 & +08 28 25 & SAB & 16.8 & 43 & 6.6\\
     NGC4536 & 12 31 53.5 & +02 27 50 & SAB & 16.8 & 67 & 6.5\\
     NGC4548 & 12 32 55.1 & +14 46 20 & SB  & 16.8 & 37 & 5.4\\
     NGC4569 & 12 34 18.7 & +13 26 18 & SAB & 16.8 & 63 & 8.3\\
     NGC4579 & 12 35 12.6 & +12 05 40 & SAB & 16.8 & 37 & 5.4\\
     NGC4654 & 12 41 25.7 & +13 23 58 & SAB & 16.8 & 52 & 4.4\\
     NGC4689 & 12 45 15.3 & +14 02 13 & SA  & 16.8 & 30 & 4.1\\
     NGC4736 & 12 48 32.4 & +41 23 28 & SA  & 4.3  & 35 &  11\\
     NGC4826 & 12 54 16.9 & +21 57 18 & SA  & 4.1  & 57 & 9.3\\
     NGC5005 & 13 08 37.6 & +37 19 25 & SAB & 21.3 & 62 & 4.8\\
     NGC5194 & 13 27 46.9 & +47 27 16 & SA pec & 7.7 & 20 & 10.5\\    
     NGC5248 & 13 35 02.4 & +09 08 23 & SAB & 22.7  & 42 & 6.3\\ 
     NGC6946 & 20 33 48.8 & +59 58 50 & SAB & 5.5 & 30 & 13.2\\ \hline
  \end{tabular}
  \end{center}

     NOTES-Units of right ascension are hours, minutes, and seconds, and
   units of declination are degrees, arcminutes, and
   arcseconds. Col.(1):Galaxy name. Col.(2)(3):Galaxy coordinates(epoch
   1950). Col.(4):Hubble Type, taken from RC2. Col.(5): Distance to the
   galaxy,  adopted from \citet{Ho97}. Col.(6):Inclination angle, given
 in \citet{Young95}. Col.(7):Projected diameter, taken from \citet{Rownd99}.
   \end{table}

\begin{table}
  \caption{CO and H$\alpha$ reference}\label{data}
  \begin{center}
    \begin{tabular}{ccccc}  \hline \hline
       & Region & CO & H$\alpha$ \\ \hline
     Global average & E & (1) & (1) \\
     central 45 arcsec &  D & (2) & (2) \\
     central 16 arcsec & C & (3) & (4) \\
     central 6  arcsec & B & (5) & (4) \\
     central 3  arcsec & A & (6),(7) & (8) \\ \hline  
     \end{tabular}
    \end{center}
   
     Notes- Reference(1):\citet{Young96} (2):\citet{Rownd99}
 (3):\citet{NN01} (4):\citet{koopman} (5)\citet{BIMA}
 (6):\citet{sakamoto99} (7):\citet{sofue03} (8):\citet{Ho97}
Data for the central 6 arcseconds were retrieved only for NGC4303,
 NGC4321, NGC4535, NGC4548 and NGC4569.  Data for the central 16
 arcseconds were  retrieved only for NGC4254, NGC4303, NGC4501, NGC4535
 and NGC4569.  Global average, central 45 and 3 arcsecond data were
 retrieved for all of the sample galaxies.  Data from (2), (4), (6) and
 (7) were traced from radial profiles given in the reference.  For
 galaxies which overlapped in (6) and (7), data from \citet{sofue03} were used.
 \end{table}

\newpage

\section{Data}

 We have made high resolution (typically 3 arcsecond)
 $\atom{CO}{}{12}$(J=1-0) observations of galaxies in the Virgo
 Cluster, with the NMA(Nobeyama Millimeter Array).  Observational
 parameters and data are presented in \citet{sofue03}.   Using the
 obtained data and other
 high resolution CO data obtained mainly at NRO (Nobeyama Radio Observatory), and further combining it with low
 resolution single dish observations in other studies to account for intermediate gas density
 regions, we have studied the validity of the Schmidt law at a wide
 density range within 23 spiral galaxies.  Fundamental parameters of
 the sample galaxies are listed in table \ref{prop}.    

\subsection{Sample selection}

\citet{Rownd99} have found that the Schmidt law holds better when
data are sampled locally within galaxies, with smaller beam size.  However, their study used CO
and H$\alpha$ data at 45 arcsecond resolution, too large to
consider high density regions which are typically small in size.  In
order to consider the Schmidt law for
a wide range of gas density, we have chosen galaxies which have been observed
in several beam sizes.  Twenty three spiral galaxies were chosen, 14 of which are
members of the Virgo Cluster.  All are spirals, SA, SAB, or SB in
morphology.  Most samples have 3 data points
corresponding to different beam sizes, expressing different density ranges.  Some galaxies have been observed
in more beam sizes.  NGC4254, NGC4321, NGC4501, NGC4548, and NGC4579 have 4 data points, and
NGC4303, NGC4535, and NGC4569 have 5 data points.  The inclination angle
$i$ (degrees), total blue magnitude $\mathrm{B_{tot}}$, and peak
$\atom{CO}{}{12}$ (J=1-0) line antenna
temperature $T_A^*$ (mK) are in the range $18 \le i \le 74$, $8.99 \le
\mathrm{B_{tot}} \le 12.08$, and $21 \le T_A^* \le 312$, respectively.

\subsection{CO data}

CO and H$\alpha$ data used in this study are listed in table \ref{data}.  Much of the
interferometric data were obtained in the course of ``Virgo
High-resolution CO Survey'', a long-term project
at Nobeyama Millimeter Array \citep{sofue03}.  We have combined
data with different spatial resolutions, so that data with low resolution, which
averages a large area within the galaxy would give low density values,
whereas high resolution data at galactic centers would give high
molecular densities.  Molecular gas densities in the literature were recalculated using a conversion
factor of $X_{\mathrm{CO}}=2.0\times 10^{20}$ ($\mathrm{cm}^{-2} \cdot
\mathrm{K}^{-1} \cdot \mathrm{km}^{-1} \cdot \mathrm{s}$) and the formulation
\begin{equation}
\bigg[ \frac{\Sigma_\mathrm{SMD}}{\mathrm{g} \cdot \mathrm{cm}^{-2}} \bigg] = 2\times \bigg[ \frac{\mathrm{m_H}}{\mathrm{g}} \bigg] \times X_{\mathrm{CO}}\bigg[ \frac{I_{\mathrm{CO}}}{\mathrm{K} \cdot \mathrm{km} \cdot \mathrm{s}^{-1}} \bigg] \times 1.6
\end{equation}
where $\mathrm{m_{\mathrm{H}}}$ and $I_{\mathrm{CO}}$ are the mass of a hydrogen atom, and CO
intensity, respectively.  $X_{\mathrm{CO}}$ is poorly understood for the
sample galaxies, and the value we adopt is the mean value of those
derived for our galaxy and other disk galaxies (e.g.,
\citet{sanders}, \citet{bloemen}, \citet{solomon}, \citet{nakai}).  The
factor 1.6 accounts for mass of heave elements.
Gas densities for each of the regions within the galaxies were derived in the following manner:

\begin{enumerate}
\item
The highest resolution data ($\sim 3$ arcseconds, see table 2) was used to derive the gas mass at
the central-most region.  We call this region $A$ 

\item
Then, the second-highest resolution data (6, 16 or 45 arcseconds, the
     highest for the individual galaxy for which we have retrieved data) which was also centered on the
     nucleus was used to derive the gas mass within the beam. 

\item
Gas mass derived from 1. was subtracted from that derived in 2.  This
     would give the gas mass confined within the annulus corresponding to
     the difference in the area of the two beam sizes.  We call this
     region in the annulus $B$.

\item 
The same procedures 1 through 3 were repeated for larger beam sizes, and then the surface gas densities were
     derived for each of the annuli.  Each of the annuli are labeled
     $C$, $D$, and $E$, respectively, in the order of increasing beam
     size.  Note that region $E$ uses the global averaged value, and is
     not actually derived from a single beam.
\end{enumerate}

\begin{figure}
  \begin{center}
   \rotatebox{0}{\FigureFile(100mm,80mm){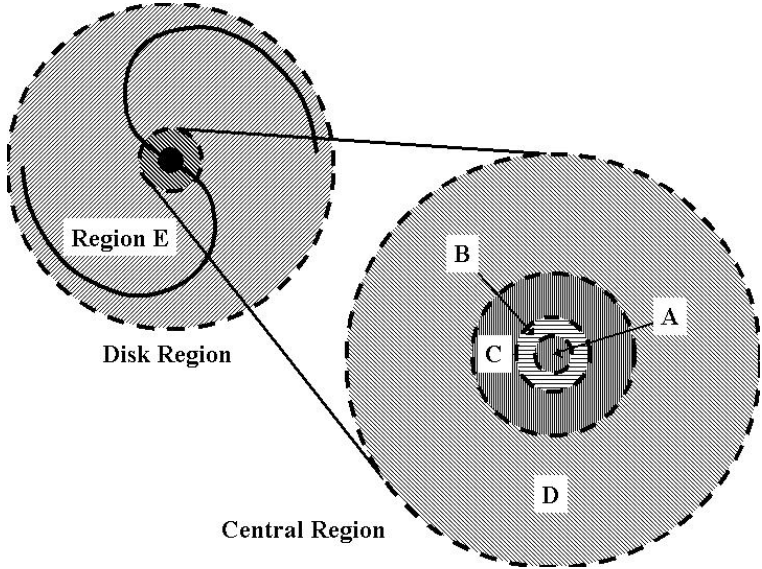}}
  \end{center}
  \caption{Schematic picture of the regions A through E.  The dotted
 lines indicate the edge of each of the regions.  The regions are scaled
 using an arbitrary global diameter (edge of region E) of 5 arcminutes
 (the average for Virgo galaxies). }\label{fig1}
\end{figure}

This procedure divides the galaxy into several annuli, so that derived
gas densities give radial dependent values.  We present a schematic
picture of the regions in figure \ref{fig1}.

\subsection{H$\alpha$ data}
The calculation of the star formation rate $\Sigma_{\mathrm{SFR}}$
$\mathrm{M_{\solar}} \mathrm{pc}^{-2} \mathrm{yr}^{-1}$ 
was conducted following \citet{K98}, using the
H$\alpha$ luminosity emitted from ionized gas around young massive
stars.
\begin{equation}
\bigg[\frac{\Sigma_{\mathrm{SFR}}}{\mathrm{M_{\solar}} \cdot \mathrm{pc}^{-2} \cdot \mathrm{yr}^{-1}}\bigg] = \bigg[\frac{L(\mathrm{H}\alpha)}{1.26 \times 10^{41} \,  \mathrm{erg} \cdot \mathrm{s}^{-1}}\bigg] \times \bigg[\frac{S}{\mathrm{pc}^{2}}\bigg]^{-1}
\end{equation}
  where $S$ is the projected area of H$\alpha$ emission.  Procedures as
  same as the CO data
(1. thorough 4.) were applied to the H$\alpha$ data for the
calculation of the $\Sigma_{\mathrm{SFR}}$. 

Galactic extinction has been corrected for all of the data.  The original
references adopt different extinction laws, and they were recalculated
using the same form of $A(H\alpha)=0.08(\csc |b|-1)$ in
magnitudes, were $b$ is
the Galactic latitude, were possible (regions A, D, and E).  In the original references (see table 2),
galactic extinction for regions B and C were corrected using Landolt standards or the standard extinction curve
at Kitt Peak National Observatory or Cerro Tololo Inter-American
Observatory.  The references for regions B and C do not list the standard stars or the data
before correction, so the corrected data from the original references were
used without adopting the same extinction law.  This will not make any
significant changes, however, because the typical extinction due to the Galaxy is
of order 0.01 to 0.1 mag; on the logarithmic scale, as we will use, this
will be 0.004 to 0.04 on the ordinate.  We therefore conclude,
that Galactic extinction for the samples are satisfactorily corrected.

 The reader should refer to the original
references (see table 2) for details.  Extinction within
the sample galaxies (internal extinction) plays a
role in the determination of the true H$\alpha$ luminosity at the
central regions of galaxies, and the effects
of extinction correction will be discussed in section 4.

\begin{figure}
  \begin{center}
    \FigureFile(100mm,100mm){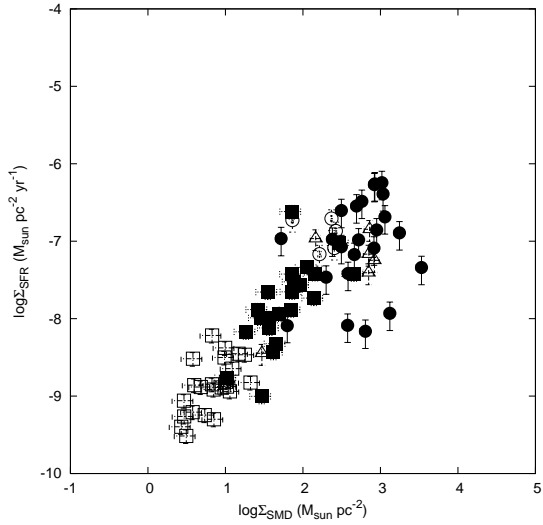}
  \end{center}
  \caption{The $\Sigma_{\mathrm{SMD}}-\Sigma_{\mathrm{SFR}}$ for all of
 the sample galaxies.  Internal H$\alpha$
 extinction has not been corrected for.  Abscissa is
 $\log \Sigma_{\mathrm{SMD}}$, ordinate $\log \Sigma_{\mathrm{SFR}}$.  Filled circles, triangles, open
 circles, filled squares, and open squares stand for regions A, B, C, D,
 and E, respectively.  Horizontal error bars for filled circles lie
 within the symbol.  Notice the several points in region A which are
 deviated from other plots.}\label{fig2}
\end{figure}

\begin{figure}
  \begin{center}
    \FigureFile(100mm,100mm){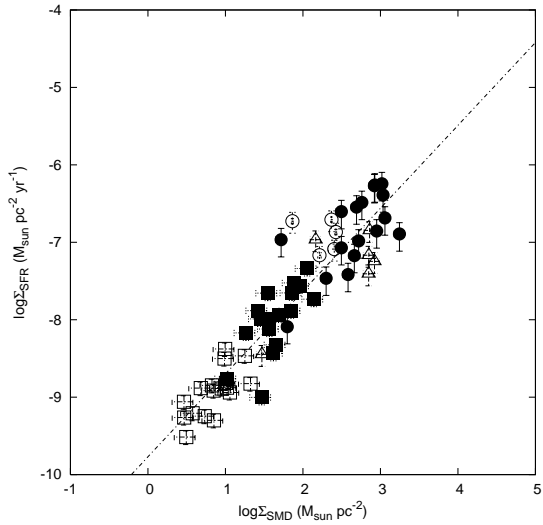}
  \end{center}
  \caption{The same figure as \ref{fig2}, with NGC2903, NGC3593, NGC4736,
 NGC4826, NGC5194, and NGC6946, excluded as mentioned in section3.  A
 best fit yields $N=1.07\pm0.06$.}\label{fig3}
\end{figure}

\begin{figure}
  \begin{center}
    \FigureFile(100mm,100mm){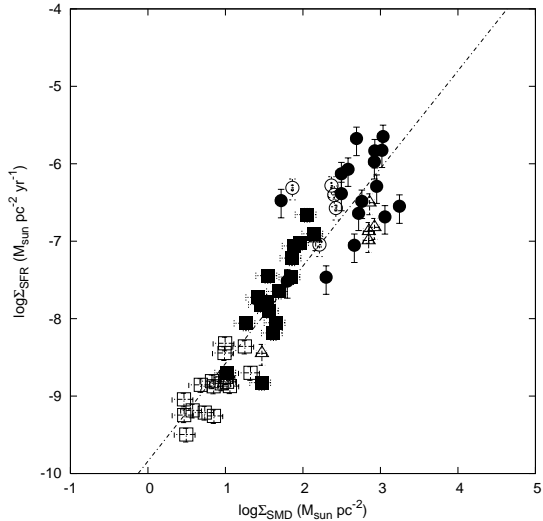}
  \end{center}
  \caption{The same figure as \ref{fig3}, corrected for H$\alpha$
 extinction.  A best fit yields $N=1.33\pm0.08$.}\label{fig4}
\end{figure}

\section{Results}

Figure \ref{fig2} shows the $\Sigma_{\mathrm{SMD}}$ versus $\Sigma_{\mathrm{SFR}}$
on a logarithmic scale, for each of the regions in the
sample galaxies.  We see clearly that data plotted with smaller
beam sizes account for the high density regime of the Schmidt law, extending to densities as high as $10^3
\mathrm{M_{\solar}} \mathrm{pc}^{-2}$.  For the highest densities, data seems to
be dispersed from the correlation. 
 However, we find that galaxies which
are dispersed from the correlation are those in the sample
whose distances are closer to our Galaxy; namely, NGC3593, NGC4736, NGC4826,
NGC6946 are only several Mpc away compared to 16.8 Mpc for most of the
sample.  The resolution of 3 arcseconds corresponds to under 100 pc for
these galaxies.  An investigation of the coordinates of the center of
these galaxies in H$\alpha$ and CO will reveal that the CO intensity
peaks are offset from H$\alpha$ centers, often by more than 3
arcseconds.  We have excluded these
galaxies in further analysis because there were possibility
that the small beam of $\sim$3 arcseconds do not include both the star
forming region and its counterpart molecular cloud which is generally
several$\times 10$ pc in size.  In general, star
forming regions do not coincide with the molecular clouds which give birth to them.
This is seen in spirals arms, where the optical and the molecular
spiral structure are offset.  If only the molecular clouds are observed
and not the star forming regions which are physically coupled to it, it
will result in the gas density to show excess from the expected correlation.
   The data points for the 4 galaxies named above, seem to have this property.  Although
not significantly different from the other galaxies, NGC2903 and NGC5194 which have
similar spatial resolution (under 100 pc) in Region A were also
excluded.  Moreover, it should be noted that H$\alpha$ luminosities of NGC4736, NGC4826, NGC5194
and NGC6946 were given only as lower limits in \citet{Ho97}.  This 
leaves a subset of 17 galaxies in our sample.  The same figure with these galaxies
excluded are shown in figure 3.  A tighter correlation can be seen
through the gas density range.     
Figure 3 clearly shows that high density regions A, B, and C, and the lower
density regions D and E, can be fitted with the same line.  A least
squares fit to figure \ref{fig3}
yields a Schmidt law index $N$ of $1.14 \pm 0.08$, allowing for errors
in both of the axes.

\section{Discussion}

\subsection{Error estimation}
\begin{flushleft}
(1)Derivation of $\Sigma_{\mathrm{SMD}}$
\end{flushleft}

  In general, we refer to gas mass as the total of both molecular and atomic
mass.  However, the legitimacy of using either or both of the hydrogen
gas is not well known.  While \citet{K89} found that total gas
density correlates better with the star formation rate, others such as
\citet{Gao04} believe that denser (hence, $\mathrm{H_2}$ rather than HI)
gas correlates better with the star formation rate.  \citet{Boissier}
 find that total and molecular gas both correlate well with
 $\Sigma_{\mathrm{SFR}}$, but the tight correlation seen in using the total
 gas is driven by the correlation of molecular gas with the star formation
 rate.  The recent trend
seems to be in favor of using only molecular gas, but it is still
controversial.  We use only molecular gas in our study, on the following
grounds; (1) massive star formation presumably occurs in molecular clouds
cores, where HI is less abundant.  On the standpoint that star formation
 is a two step process where $\mathrm{H_2}$ clouds are first formed from HI, and 
 stars are then formed from $\mathrm{H_2}$, we conclude that $\mathrm{H_2}$ should be a more direct
 tracer of the star formation rate.  (2) We are interested in high density
regions of galaxies, and have used galactic centers.
These regions are known to be HI deficient, so even if HI plays a role in
 star formation, its mass should be negligible.

Another uncertainty resides in the CO to $\mathrm{H_2}$
conversion factor $\mathrm{X_{CO}}$.  The constant value of
$\mathrm{X_{CO}}=2.0\times 10^{20}$ is highly questionable.  However, the conversion
factor for each of the sample galaxies is not known, and moreover it's variation within
those galaxies are potentially difficult to know; we have used the
average value of our galaxy and galaxies with known $\mathrm{X_{CO}}$ 
uniformly for all of the data for the sake of simplicity.
\citet{Boissier} have found that by using
 metallicity dependent $\mathrm{X_{CO}}$,
$\Sigma_{\mathrm{SMD}}$ should change typically by a factor of 2 to 3 in the
central regions compared to disk regions of nearby spirals, in the sense
of decreasing gas mass.  This, however, does not make disastrous changes compared to using a constant
conversion factor, because 
factors of 2 to 3 makes little difference to the positions in which data
are plotted on the log$\Sigma_{\mathrm{SMD}}$ -
log$\Sigma_{\mathrm{SFR}}$ scale.  Again, we adopt the constant
$\mathrm{X_{CO}}$ for the sake of uniformity and simplicity.

 The highest resolution measurements use interferometers, which has the
 ``missing flux'' problem.  Spatially extended sources can not be detected, which
 has the effect of underestimating the surface density.  Figure 3 indicates that compared to region D with 45
 arcsecond resolution, region A and B with the interferometric
 observations are typically an order of magnitude higher in
 $\Sigma_{\mathrm{SMD}}$.  Even compared to region C with 16 arcsecond
 resolution, region A is higher typically by a factor of about 3.  This
 indicates that the missing flux has no significant effect on $\Sigma_{\mathrm{SMD}}$.

The horizontal error bars in figure 3 do not include the uncertainties
given above.  They are typically 15 to
30\%, and are mainly due to
errors in flux calibration, given in the original references.

\begin{flushleft}    
(2)H$\alpha$ extinction
\end{flushleft}

The H$\alpha$ emission lines are subject to extinction within the target galaxies, and we must also account for
this in the analysis.  A global, uniform extinction correction of $\sim$ 1 mag. is
suggested by \citet{kenken83}, but this is an oversimplification, because
central regions of 
galaxies are generally thought to be subject
to larger extinction than disk regions.  Extinction for Region A can be
calculated from E(B-V) magnitudes given in \citet{Ho97}, but extinction
for the other regions B through E are not known.  

An extinction model where
the extinction increases in proportion with the hydrogen column number
density $\mathrm{N_H}$ may be used for these regions.  Using
$\mathrm{N_H}=2\times 10^{21}\mathrm{A_V}$ (\citet{schultz}, \citet{bohlin}, \citet{sneden},
\citet{fazio})and $\mathrm{N_H}=\mathrm{X_{CO}}I_{\mathrm{CO}}$,
 the H$\alpha$ extinction $\mathrm{A_R}$ can be calculated
by $\mathrm{A_R}=0.75\mathrm{A_V}$ (\citet{rieke}).  We assume that $\mathrm{N_H}$ is
dominated by molecular hydrogen, because HI and $\mathrm{H_2}$ are
generally spatially offset in galaxies disks, divided by a transition
region.  As long as we calculate $\Sigma_{\mathrm{SMD}}$ using only
$\mathrm{H_2}$, HI should only play a minor role in the extinction, even
when calculating global values in Region E.  However, this extinction
model is known to yield overestimates in regions of higher density
\citep{wong02}, and will give unrealistic values of $\mathrm{A_R}$ magnitude up
to 5 mag. in Region C, 10 mag. in Region B, and even more in Region A.
This is clearly an overestimation, because the gas consumption timescale
$\tau$ derived for these regions using this correction (see section 4.4) will yield $\tau$
of only several hundred years, too short to explain the many number of
galaxies with such star forming central regions.  Therefore, we adopt the
same extinction derived in Region A, for Regions B, and C.  For Regions
D and E, the extinction model explained above is adopted.  A source of
uncertainty in applying extinction based on molecular density for
regions D and E, are the difference in resolution for CO and H$\alpha$
measurements.  Light from HII regions is absorbed by dust associated
with molecules that are
directly between the observer and the HII region in the line of sight.
 However, the observing beam of the CO observation is far larger than
that of the H$\alpha$ observation.  Therefore, we can only apply
extinction correction based on the average gas density of the region
within this CO beam size, whereas the actual extinction of H$\alpha$
light occurs at a smaller scale.  This may cause the extinction
correction to be an underestimate, but the true correction is impossible unless
the H$\alpha$ emitting region is completely resolved in the CO observation.  In the
case of regions D and E where the beam size is 45 arcseconds, this
effect may not be negligible.  However, because existing CO
observations cannot resolve individual HII regions, we will not further
refer to this problem.  The effect of neglecting this possible
underestimation will be discussed briefly in section 4.2.

Figure \ref{fig4} is the same as figure \ref{fig3}, with the extinction corrected as explained
above.  A least squares fit yields $N=1.33\pm0.09$, higher than
$N=1.14\pm0.08$ for no extinction correction.  Even with the extinction
corrected, however, $N$ is lower than the widely accepted
value of 1.4 given by \citet{K98} which includes IR luminous starbursts at dense gas regimes. 

Other errors are as follows:
H$\alpha$ data used in region D been traced from figure 2 in
\citet{Rownd99}, and regions B and C from figure 5 in \citet{koopman}.  We
assign an error of 0.1 on the ordinate of these data.  Other errors are
mainly due to difficulty in continuum subtraction. See original paper for details.

\subsection{$\Sigma_\mathrm{SMD}-\Sigma_\mathrm{SFR}$ relation}
 A least-squares fit to figure \ref{fig4} results in
$$\log\Sigma_{\mathrm{SFR}}=(1.33\pm0.09) \log \Sigma_{\mathrm{SMD}}-9.95\pm0.2,$$ and a
best fit to the same data using only region E, or the disk, results in
$$\log\Sigma_{\mathrm{SFR}}=(1.51\pm 0.6) \log\Sigma_{\mathrm{SMD}}-10.19\pm0.4$$
Similarly, a best fit using only regions A, B, C and D, or the central
regions, will yield $$\log\Sigma_{\mathrm{SFR}}=(1.22\pm0.15) \log \Sigma_{\mathrm{SMD}}-9.68\pm0.4$$
The index and the constant of the fits for the disk and central regions
are both well within the range of error, although errors for each of the
regions are large.  
This is consistent with the idea that the data plots in the central
regions have the same Schmidt law as in the disk regions.

The beam size of the highest density region A (typically 3 arcseconds)
corresponds to a projected scale of 250 pc for galaxies in the
Virgo cluster.  The Schmidt law is thought to be an empirical law which holds at kpc
scales within galaxies, but our results imply that smaller scales of
several hundred parsecs may still be valid in considering the Schmidt
law.  It may be noted, that regions close to the nucleus are subject to
physical conditions different from the disk.  Besides the dense gas
tracers such as HCN and CO(J=2-1) observed in these regions, rotation curves in the
central kpc are also
known to be different from disks, in that rotation becomes nearly rigid
compared to nearly flat in the disks \citep{sofue03b}.  This will lessen the effect of
shear which work on the clouds, and may change the efficiency of star
formation based on gravitational collapse.  These effects could work as
systematic change in the relation between gas content and star formation
rate in the central regions.  However,   
the index $N$ of the Schmidt law which we have derived from simply
fitting figure \ref{fig4} with a least squares fit indicates no
systematic difference in the Schmidt law within the range of error.
  We conclude that the validity
of the Schmidt law at densities from $10^0 \mathrm{M_{\solar}} \mathrm{pc}^{-2}$ up to $10^3 \mathrm{M_{\solar}} \mathrm{pc}^{-2}$
does not change.

\begin{figure}
  \begin{center}
    \FigureFile(100mm,100mm){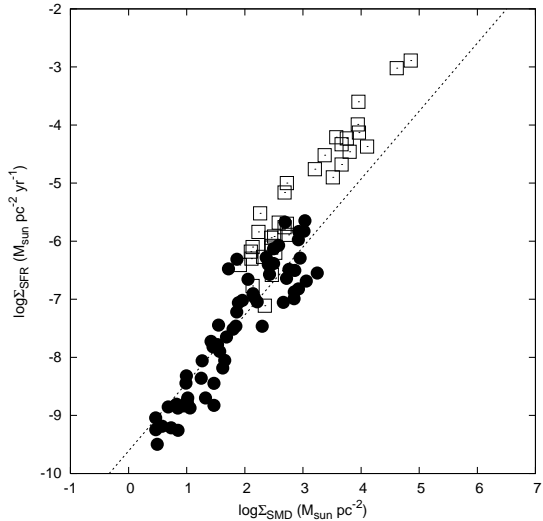}
  \end{center}
  \caption{The composite Schmidt law.  Filled circles are the normal
 galaxies in this study, and open squares are the IR luminous
 starbursts from \citet{K98}.  The
 line is a best fit only to the normal galaxies.  Notice that the
 starbursts are systematically offset from the Schmidt law of normal
 galaxies in the sense of higher $\Sigma_{\mathrm{SFR}}$.  A best fit
 for both of the data together will yield
 $\log\Sigma_{\mathrm{SFR}}=(1.41\pm0.07) \log\Sigma_{\mathrm{SMD}}-9.83\pm0.18$}\label{fig5}
\end{figure}

 It is difficult to consider the Schmidt law for normal galaxies at even
 higher densities, because doing this will require that we observe with even
 better spatial resolution; which in turn, will give rise to the problem
 that we may not be able to observe the molecular cloud and its
 counterpart star forming region within one beam, as we have seen in
 section 3 with galaxies at several Mpc from us.  As can be understood
 from this example, the Schmidt law is fundamentally a correlation seen
 when averaged over a considerable scale.  It may be said that this
 linear correlation breaks down when averaged over an area of under
 $\sim 10^2$ pcs.  Thus, for higher
 densities, we must use IR luminous galaxies which have unusually
 high molecular densities like the studies by \citet{K98}.  A
 superposition with these IR luminous starbursts from \citet{K98} and
 our study is shown in figure \ref{fig5}.  Error bars are not shown, but
 are the same as figure 4.  The high density end of
 normal galaxies exhibit similar $\Sigma_{\mathrm{SMD}}$ and
 $\Sigma_{\mathrm{SFR}}$ as the lower density end of the starbursts, but the starbursts display
 higher $\Sigma_{\mathrm{SFR}}$ compared to the Schmidt law derived for
 normal galaxies at densities above $\Sigma_{\mathrm{SMD}}=10^3 \mathrm{M_{\solar}}
 \mathrm{pc}^{-2}$.  It is important to bear in mind, though, that
 $\Sigma_{\mathrm{SFR}}$ of the IR starbursts have been derived from
 IR luminosities, possibly introducing inconsistencies between
 H$\alpha$ derived rates.  $\Sigma_{\mathrm{SFR}}$ derived from
 H$\alpha$ will indeed give lower $\Sigma_{\mathrm{SFR}}$ if extinction
 is not corrected.  However, we have corrected for H$\alpha$ extinction
 in our analysis, and \citet{Kewley} find that star formation rates
 derived from extinction corrected
 H$\alpha$ and FIR luminosities agree well.  Therefore, this systematic difference between starbursts and normal
 galaxies we see in figure \ref{fig5} may indeed be real, implying that again, we
 must establish the relation between gas and star formation rates for normal galaxies first in order to
 gain a fundamental knowledge of the relation between gas content and
 star formation rate.  A least squares fit to the normal and starbursts
 together, will give $N=1.41\pm0.07$, consistent with the ``composite''
 Schmidt law derived by Kennicutt(1998). 

   It is justified to say that the derived slope of $N=1.33$ is
    relatively reliable, despite the unquantifiable uncertainties, such as
    $\mathrm{X_{CO}}$, the missing flux problem stated in section
    4.1, and the problem of underestimating H$\alpha$ extinction stated
    in 4.2.  This is because allowing for a systematic
   decrease in $\mathrm{X_{CO}}$ by a factor of 3 with galactocentric radius will
   raise the slope, while on the other hand, correcting for the missing
   flux and underestimated extinction will increase the density slightly
    in regions A and B and the star formation rate in regions D and E resulting in the
   decrease of the slope.  These effects will offset to some extent, and changes
   in the slope will not be so significant.  Comparison with Starbursts
    and normal galaxies will not be affected significantly, because results by
   Kennicutt(1998) will also have to be corrected in accordance.  

\subsection{Interpretation of the Schmidt law index $N$}    
The widely accepted value of $N=1.4$ by \citep{K98} is unchallenged
regarding its density range of order $\sim$5, and a simple
interpretation of this index is often given that the star formation rate
volume density
$\rho_{\mathrm{SF}}$ scales with gas volume density
$\rho_{\mathrm{gas}}$ divided by the free-fall timescale of the gas
cloud, $t_{ff} \propto (\sqrt{{\mathrm{G} \rho_{\mathrm{gas}}}})^{-1}$.
This will give
\begin{equation}
\rho_{\mathrm{SF}} \propto \frac{\rho_{\mathrm{gas}}}{{(\sqrt{\mathrm{G} \rho_{\mathrm{gas}}})}^{-1}} \propto {\rho_{\mathrm{gas}}}^{1.5},
\end{equation}
which may explain the value of $N=1.4$.  However, this explanation assumes that we
are observing a single gas cloud as it gravitationally collapses and
eventually forms a star.  In actual observations, we are observing an
area of the galaxy with a beam in scales of several hundred parsecs.
Naturally, it is not possible to observe a single cloud, but instead the
observer is averaging the
gas density and star formation rate of many clouds which are typically
several$\times$10 parsecs in linear scale.  In such scales, the increase in the gas
density $\rho_{\mathrm{gas}}$ (or, $\Sigma_{\mathrm{SMD}}$) should not be interpreted as an increase in the
density of the individual clouds.  Instead, we should regard the change as an
increase in the number of molecular clouds, within that beam.  Therefore,
as long as the observing beam is much larger than the typical size of a single
molecular cloud, the $\Sigma_{\mathrm{SFR}}$ should increase in proportion with
$\Sigma_{\mathrm{SMD}}$, giving $N=1$ as the natural result. 
A best fit to the IR starbursts from \citet{K98} will give
$N=1.28\pm0.08$, similar to that for the normal galaxies in
our study.  However, the factor $A$ of the Schmidt law (provided $\Sigma_\mathrm{SFR}=A\Sigma_\mathrm{SMD}^{N}$), determined by the y-axis intercept
of figure \ref{fig5}, is higher for these starbursts compared to the normal
galaxies, as can be seen clearly from figure \ref{fig5} with
$\log{A}=-9.09\pm0.24$.  The datasets for the starbursts have been
recalculated using the same $\mathrm{X_{CO}}$ as this study.  The transition in the
Schmidt law from normal galaxies to starbursts seems to be
step-like, or that there are two systems in the Schmidt law, with
systematically different star formation rates.  

\begin{table}
  \caption{The Schmidt Law index}\label{index}
  \begin{center}
    \begin{tabular}{lccc}  \hline \hline
     Region & H$\alpha$ extinction correction & N &  $\log \mathrm{A}$  \\ \hline\  
      All   &  No     &    $1.14\pm0.08$  & $-9.87\pm0.20$ \\ 
      All   &  Yes    &    $1.33\pm0.09$  & $-9.95\pm0.20$ \\
      E   &  Yes    &    $1.51\pm0.60$  & $-10.19\pm0.4$ \\
      A+B+C+D &  Yes    &    $1.22\pm0.15$  & $-9.68\pm0.4$ \\  
      Kennicutt(1998) & Yes & $1.31\pm0.04 \, (1.4\pm0.15)$ &
     $-9.45\pm0.07 \, (-9.61\pm0.13)$ \\
     Starbursts & No (IR) & $1.28\pm0.08$ & $-9.09\pm0.24$ \\ \hline
     \end{tabular}
    \end{center}
    The Schmidt Law index and the constant for various regions in our study, and
 that derived by Kennicutt(1998).  A constant value of 1.1 mag. was
 adopted for H$\alpha$ extinction in Kennicutt(1998).  The values for
 Kennicutt(1998) has been changed to include only molecular gas, whereas
 the values in parentheses are from the original paper.
    \end{table}

\subsection{The Star Formation Efficiency}
\begin{figure}
  \begin{center}
  \begin{tabular}{cc}
    \FigureFile(40mm,40mm){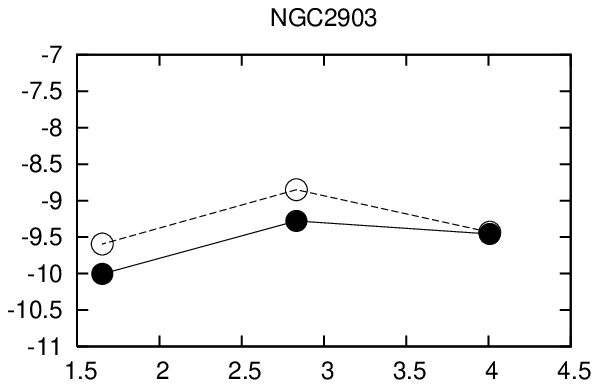}
    \FigureFile(40mm,40mm){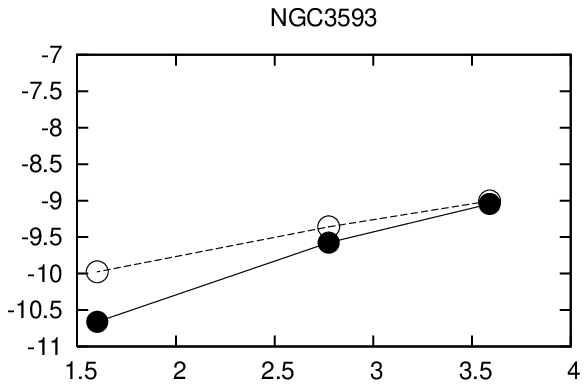}
    \FigureFile(40mm,40mm){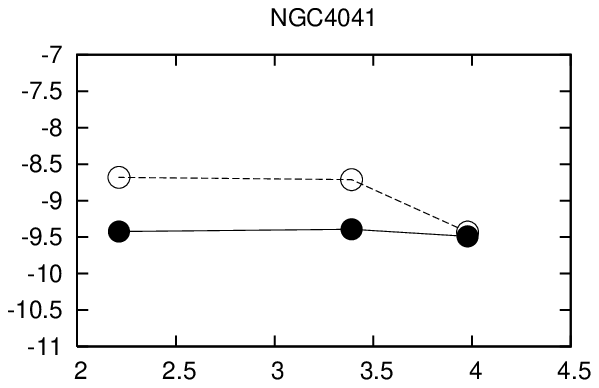}
    \FigureFile(40mm,40mm){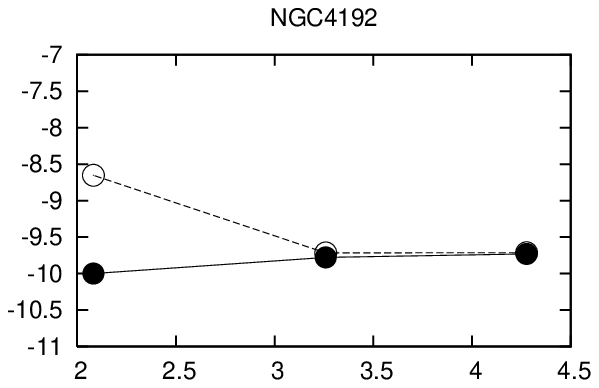} \\
    \FigureFile(40mm,40mm){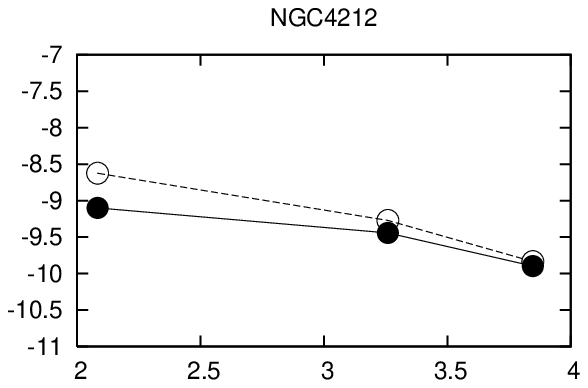}
    \FigureFile(40mm,40mm){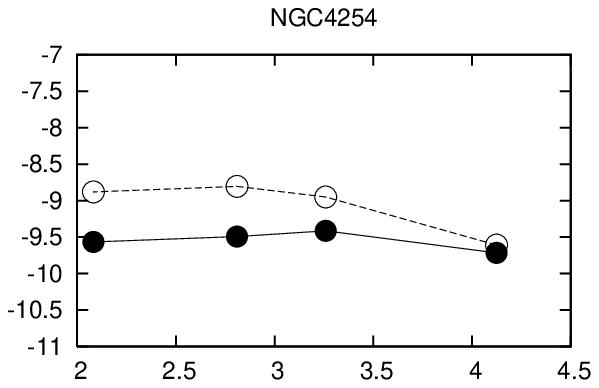}
    \FigureFile(40mm,40mm){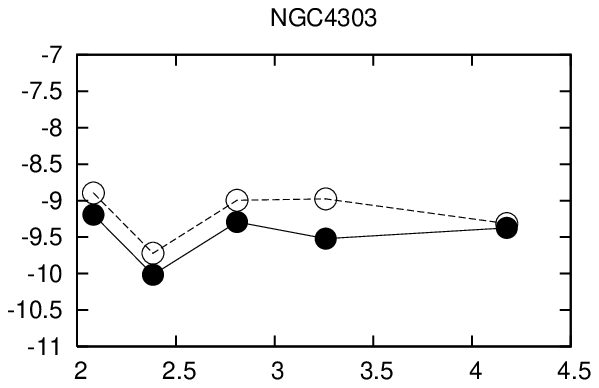}
    \FigureFile(40mm,40mm){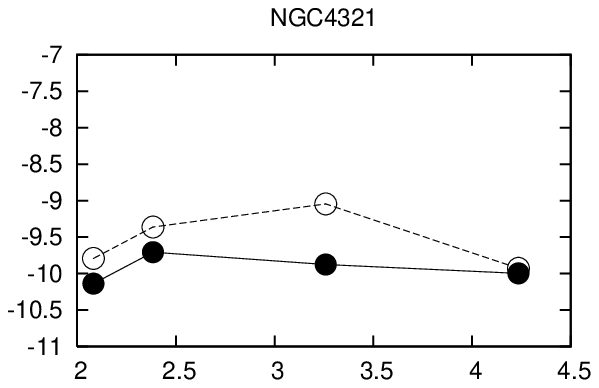} \\
    \FigureFile(40mm,40mm){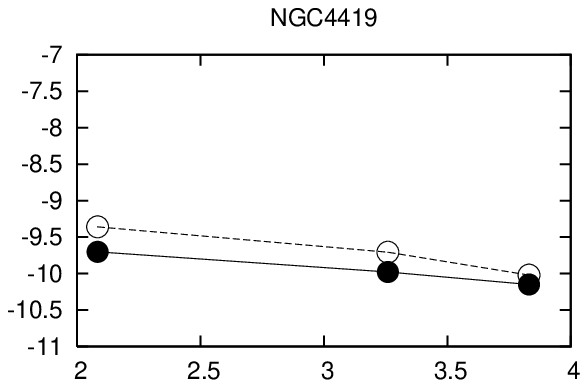}
    \FigureFile(40mm,40mm){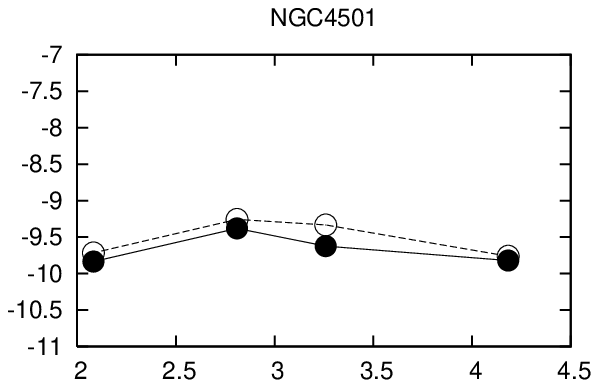}
    \FigureFile(40mm,40mm){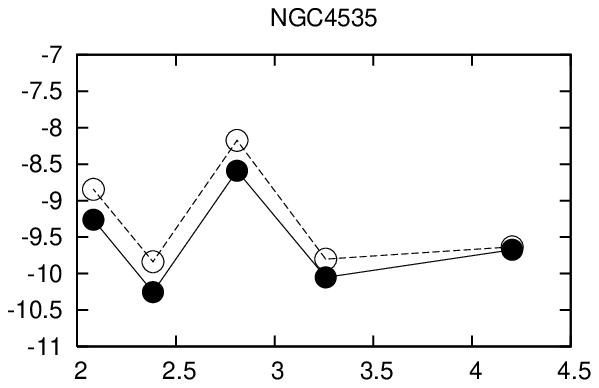}
    \FigureFile(40mm,40mm){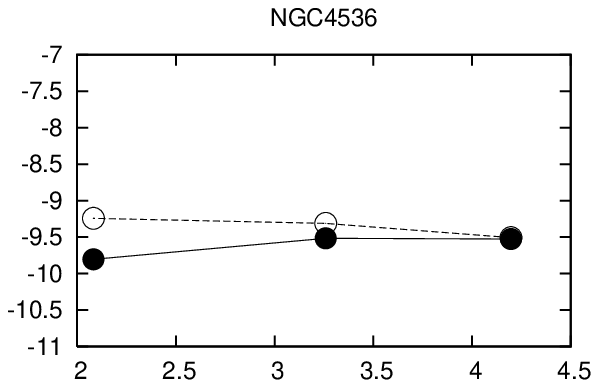} \\
    \FigureFile(40mm,40mm){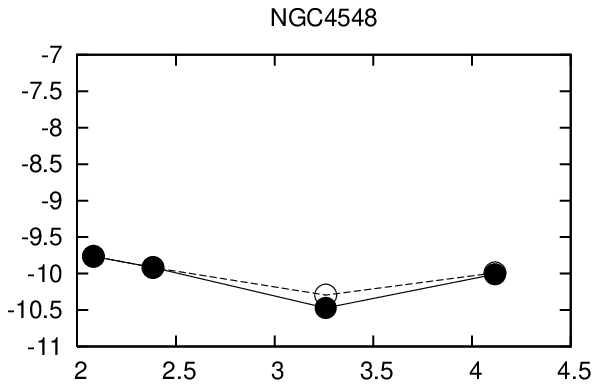}
    \FigureFile(40mm,40mm){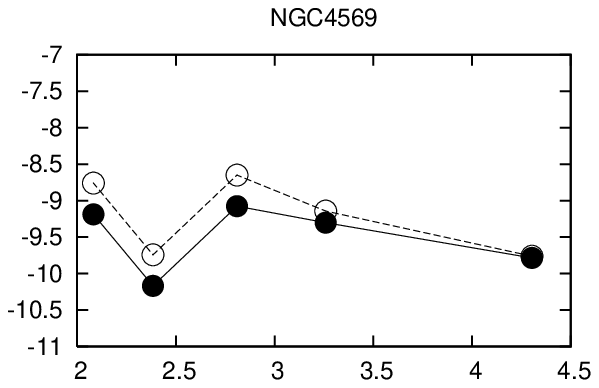}
    \FigureFile(40mm,40mm){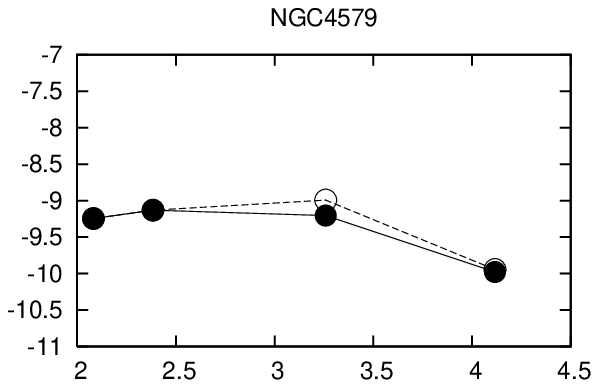}
    \FigureFile(40mm,40mm){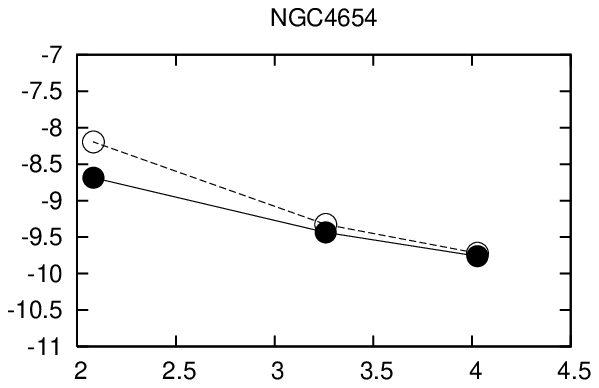} \\
    \FigureFile(40mm,40mm){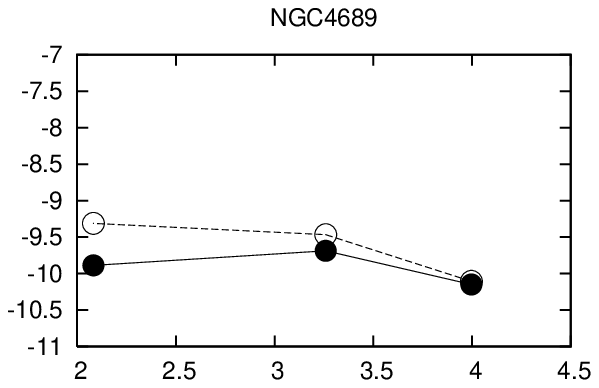}
    \FigureFile(40mm,40mm){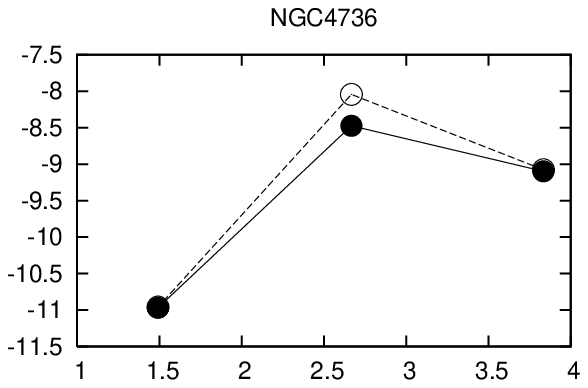}
    \FigureFile(40mm,40mm){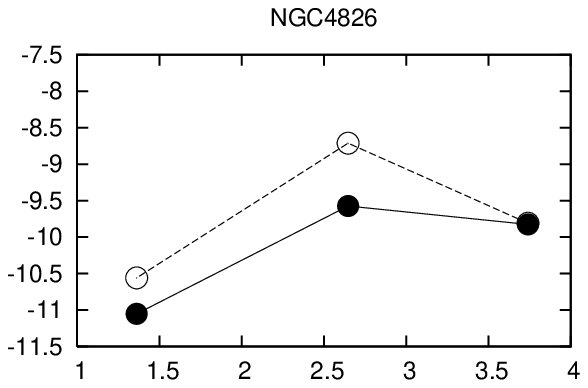}
    \FigureFile(40mm,40mm){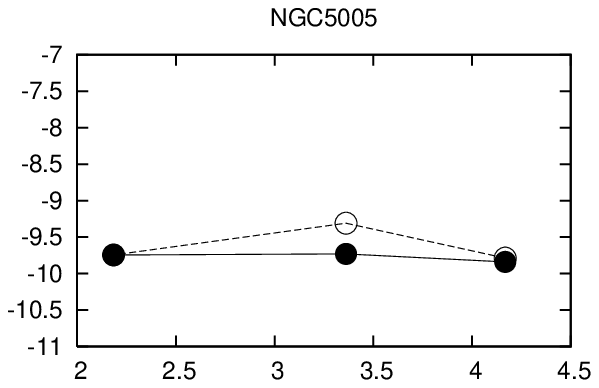} \\
    \FigureFile(40mm,40mm){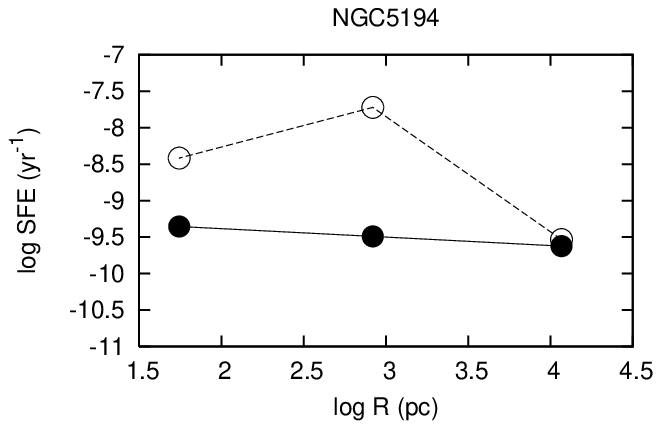}
    \FigureFile(40mm,40mm){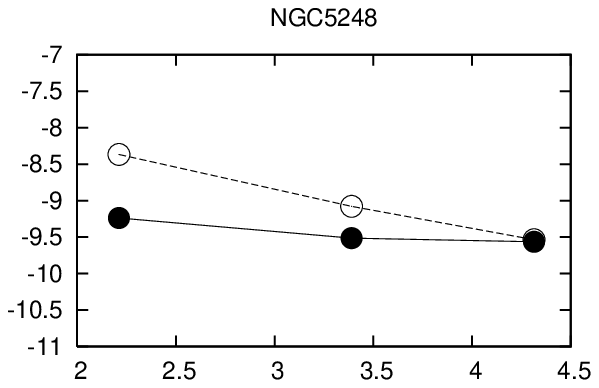}
    \FigureFile(40mm,40mm){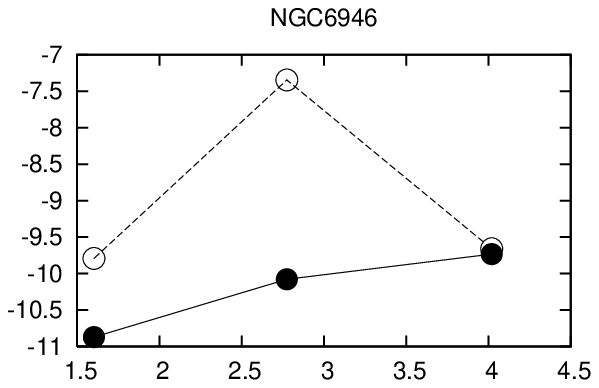}
    \FigureFile(40mm,40mm){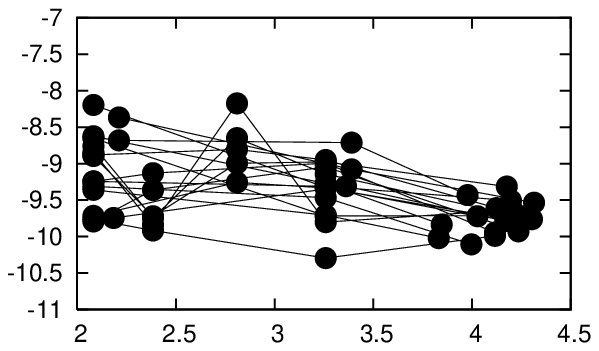}
    
     \end{tabular}
     \caption{Star formation efficiency (SFE) )vs. galactocentric radius,
   for
   all of the sample galaxies.  Filled circles are data with no
   H$\alpha$ extinction correction, and open circles are extinction
   corrected data following the method shown in section 4.  The bottom
   right figure is a superposition of the subset of 17 sample galaxies,
   all corrected for extinction.}\label{fig6}
     \end{center}
\end{figure}

  It is convenient to define a star formation
  efficiency(SFE) by
\begin{equation}
  \mathrm{SFE} \,\,\,\,\, (\mathrm{yr}^{-1}) =\frac{\Sigma_{\mathrm{SFR}}}{\Sigma_{\mathrm{SMD}}},
\end{equation}
 indicating the efficiency of gas consumption for the
  formation of stars.  \citet{Rownd99} has shown that SFE is roughly
  constant within galaxies, using 45 arcsecond resolution data.  We show
  in figure \ref{fig6} the change in SFE according to the distance from the center, using
  data in regions A through E.  The radii for the plots are taken at the
  outer edge of each of the regions.  Notice that the last panel in
 figure \ref{fig6}, with all of the galaxies superposed except those excluded in
  section 3, exhibit a roughly constant SFE of around $10^{-9}$ to $10^{-10} \, \, (\mathrm{yr}^{-1})$even in the central
  regions, with an exception of several galaxies with enhanced SFE.
  Constant SFE is a manifestation of the Schmidt law with an index
  $N=1$, and shows that the gas consumption timescale $\tau$, or the inverse of
  SFE, is about $1$ to $10$ Gyr through the entire disk and nucleus for most of the
  galaxies, and down to $0.1$ Gyr for the galaxies with enhanced SFE
  mentioned above.    

\section{Conclusions}

We have considered the relation between gas content and star formation
rate at high molecular densities in the central regions of 23 nearby
normal spiral galaxies, using recent high resolution CO observation data
and combining it with previous single dish observations.  Our main
results are as follows;

(1)The Schmidt law, or a simple power law of $\Sigma_{\mathrm{SFR}}
   \propto \Sigma_{\mathrm{SMD}}^N$, is found to be valid at densities
    up to $\Sigma_{\mathrm{SMD}}\sim 10^3 (\mathrm{M}_{\solar} \,
   \mathrm{pc}^{-2})$ for normal spiral galaxies.  Regardless of the radical physical differences
   which these high density regions are subject to, the Schmidt law
   index is found to be $N=1.33\pm 0.09$, with $N\sim 1$ in
   both central and disk regions.

(2)IR luminous starbursts, which Kennicutt (1998) used to define a
   composite Schmidt law, is found to display a systematic difference
   compared to normal galaxies.  These starbursts show
   systematically higher star formation rates compared to that expected
   from the Schmidt law for normal spirals.

(3)The star formation efficiency (SFE) is found to be roughly constant
   for more than half of our samples,
   consistent with the result by Rownd \& Young(1999), even in the
   central few hundred parsecs of spirals.  Several other galaxies
   show higher SFE in the central regions.

Further analysis of the Schmidt law, in order to gain insights in to the
physics of star formation depending on local physical states, should
involve a galaxy sampled at multiple points within its disk, categorized
by its characteristic physical conditions.  A fundamental difficulty
is that the Schmidt law has a time-averaged nature. Spatial
resolution, velocity fields,
and the local star formation timescale will alter the relation between
gas and star formation tracers.  If these uncertainties are
removed, we may see sequential differences in the observed Schmidt law
according to physical conditions.
    
Although our results imply that starbursts display a different sequence
of gas-star
formation rate relation compared to normal spirals, there are still
uncertainties in the consistency between $\Sigma_{\mathrm{SFR}}$ derived from
IR and H$\alpha$ luminosities which are difficult to quantify.  This is
because the rate of gas clouds or dust covering a newly born OB star could alter
the ``true'' star formation rate when converted into IR or H$\alpha$
luminosities. 
   In order to compare normal and starburst galaxies in
a more quantitative way, it will be important to use a uniform tracer of
star formation.

The authors are grateful to T.Handa, for his enlightening comments on
the Schmidt law index.  H.Nakanishi and S.Onodera were financially
supported by a Research Fellowship from the Japan Society for the Promotion
of Science for Young Scientists.

\begin{thebibliography}{}
\bibitem[Bloemen et al.(1986)]{bloemen} Bloemen, J.~B.~G.~M., 
et al.\ 1986, \aap, 154, 25 

\bibitem[Bohlin et al.(1978)]{bohlin} Bohlin, R.~C., Savage, 
B.~D., \& Drake, J.~F.\ 1978, \apj, 224, 132 

\bibitem[Boissier et al.(2003)]{Boissier} Boissier, S., 
Prantzos, N., Boselli, A., \& Gavazzi, G.\ 2003, \mnras, 346, 1215 

\bibitem[Cardelli et al.(1989)]{CCM} Cardelli, J.~A., 
Clayton, G.~C., \& Mathis, J.~S.\ 1989, \apj, 345, 245 

\bibitem[Gao \& Solomon(2004)]{Gao04} Gao, Y., \& Solomon, 
P.~M.\ 2004, \apj, 606, 271 

\bibitem[Helfer et al.(2003)]{BIMA} Helfer, T.~T., Thornley, 
M.~D., Regan, M.~W., Wong, T., Sheth, K., Vogel, S.~N., Blitz, L., \& Bock, 
D.~C.-J.\ 2003, \apjs, 145, 259 

\bibitem[Ho et al.(1997)]{Ho97} Ho, L.~C., Filippenko, 
A.~V., \& Sargent, W.~L.~W.\ 1997, \apjs, 112, 315 

\bibitem[Kennicutt \& Kent(1983)]{kenken83} Kennicutt, R.~C., \& 
Kent, S.~M.\ 1983, \aj, 88, 1094

\bibitem[Kennicutt(1989)]{K89} Kennicutt, R.~C.\ 1989, 
\apj, 344, 685

\bibitem[Kennicutt(1998)]{K98} Kennicutt, R.~C.\ 1998, 
\apj, 498, 541

\bibitem[Kent et al.(1991)]{fazio} Kent, S.~M., Dame, T.~M., 
\& Fazio, G.\ 1991, \apj, 378, 131

\bibitem[Kewley et al.(2002)]{Kewley} Kewley, L.~J., Geller, 
M.~J., Jansen, R.~A., \& Dopita, M.~A.\ 2002, \aj, 124, 3135 

\bibitem[Koopmann et al.(2001)]{koopman} Koopmann, R.~A., 
Kenney, J.~D.~P., \& Young, J.\ 2001, \apjs, 135, 125 

\bibitem[Nakai \& Kuno(1995)]{nakai} Nakai, N., \& Kuno, N.\ 
1995, \pasj, 47, 761 

\bibitem[Nishiyama \& Nakai(2001)]{NN01} Nishiyama, K., \& 
Nakai, N.\ 2001, \pasj, 53, 713

\bibitem[Rieke \& Lebofsky(1985)]{rieke} Rieke, G.~H., \& 
Lebofsky, M.~J.\ 1985, \apj, 288, 618 

\bibitem[Rownd \& Young(1999)]{Rownd99} Rownd, B.~K., \& Young, 
J.~S.\ 1999, \aj, 118, 670 

\bibitem[Sakamoto et al.(1999)]{sakamoto99} Sakamoto, K., Okumura, 
S.~K., Ishizuki, S., \& Scoville, N.~Z.\ 1999, \apjs, 124, 403 

\bibitem[Sanders et al.(1984)]{sanders} Sanders, D.~B., 
Solomon, P.~M., \& Scoville, N.~Z.\ 1984, \apj, 276, 182 

\bibitem[Schmidt(1959)]{schmidt} Schmidt, M.\ 1959, \apj, 129, 
243 

\bibitem[Schultz \& Wiemer(1975)]{schultz} Schultz, G.~V., \& 
Wiemer, W.\ 1975, \aap, 43, 133  

\bibitem[Sneden et al.(1978)]{sneden} Sneden, C., Gehrz, 
R.~D., Hackwell, J.~A., York, D.~G., \& Snow, T.~P.\ 1978, \apj, 223,
		       168 

\bibitem[Solomon et al.(1987)]{solomon} Solomon, P.~M., Rivolo, 
A.~R., Barrett, J., \& Yahil, A.\ 1987, \apj, 319, 730 

\bibitem[Sofue et al.(2003a)]{sofue03} Sofue, Y., Koda, J., 
Nakanishi, H., Onodera, S., Kohno, K., Tomita, A., \& Okumura, S.~K.\ 2003, 
\pasj, 55

\bibitem[Sofue et al.(2003b)]{sofue03b} Sofue, Y., Koda, J., 
Nakanishi, H., \& Onodera, S.\ 2003, \pasj, 55, 59 

\bibitem[Young et al.(1995)]{Young95} Young, J.~S., et al.\ 
1995, \apjs, 98, 219

\bibitem[Young et al.(1996)]{Young96} Young, J.~S., Allen, L., 
Kenney, J.~D.~P., Lesser, A., \& Rownd, B.\ 1996, \aj, 112, 1903 

\bibitem[Wong \& Blitz(2002)]{wong02} Wong, T., \& Blitz, L.\ 
2002, \apj, 569, 157 
\end{thebibliography}
\end{document}